\documentclass{elsart}

\usepackage{graphicx}
\usepackage{epsfig,rotating}
\usepackage[dvips]{color}

\def\be{\begin{equation}}
\def\ee{\end{equation}}
\def\ba{\begin{eqnarray}}
\def\ea{\end{eqnarray}}

\def\lsim{\raise0.3ex\hbox{$\;<$\kern-0.75em\raise-1.1ex\hbox{$\sim\;$}}}
\def\gsim{\raise0.3ex\hbox{$\;>$\kern-0.75em\raise-1.1ex\hbox{$\sim\;$}}}

\def\theta{\vartheta}

\begin{document}

\begin{frontmatter}

\title{GZK Photons in the Minimal Ultra High Energy Cosmic Rays Model}

\author[LA]{Graciela Gelmini},
\author[INR]{Oleg Kalashev} and
\author[APC,INR]{Dmitry V. Semikoz}
\address[LA]{ Department of Physics and Astronomy, UCLA, Los Angeles,
CA 90095-1547, USA.}
\address[INR]{INR RAS, 60th October Anniversary pr. 7a, 117312 Moscow, Russia.}  
\address[APC]{APC, 10, rue Alice Domon et Leonie Duquet, Paris 75205, France.}

\date{}

\begin{abstract}

In a recently proposed model the cosmic rays  spectrum at energies
 above $10^{18}$ eV can be fitted with a minimal number of unknown
 parameters assuming that the extragalactic cosmic rays are only
 protons with a power law source spectrum $\sim E^{-\alpha}$ and
 $\alpha\simeq 2.6$~\cite{dip}.
 Within this minimal model, after fitting the observed HiRes spectrum with
four parameters (proton injection spectrum power law index and maximum
energy, minimum distance to sources and evolution parameter) we compute
the flux of ultra-high energy photons due to photon-pion production, the
GZK photons, for several radio background models and average extragalactic
magnetic fields with amplitude between  $10^{-11}$ G and $10^{-9}$ G. We
find the photon fraction to be between $10^{-4}$ and $10^{-3}$ in cosmic rays at
energies above $10^{19}$ eV. These small fluxes could only be detected in
future experiments like Auger North plus South and EUSO.

\begin{small}
PACS: 98.70.Sa 
\end{small}

\end{abstract}

\maketitle

\end{frontmatter}

\section{Introduction}

The sources and the composition of  the Ultra High Energy Cosmic Rays
(UHECR), namely the cosmic rays with energy $E > 10^{18}$ eV,  are
still unknown. The highest energy cosmic rays, above $\sim 1\times
10^{19}$ eV, are likely of extragalactic origin, since they could not
be confined by the galactic magnetic fields. Low energy cosmic rays
originate within our galaxy. Two different proposals have been made for
the possible transition from galactic to extra-galactic cosmic rays in
the spectrum. Historically the ``ankle", a feature close to  $1 \times
10^{19}$~eV, was interpreted as the transition from a rapidly falling
galactic flux component to a flatter spectrum of extragalactic
originsubdominant at lower energies. Alternatively,  the ``ankle"
feature can be interpreted as an absorption ``dip" at energies
$E=3-10$~EeV~\cite{dip}, due to the
propagation of extragalactic protons over large distances in the
cosmic microwave background (CMB)~\cite{dip-old}. The transition from
a galactic to extragalactic cosmic rays  would then happen at lower
energies, $\sim 1 \times 10^{18}$~eV or below. This would agree with
the indication of a transition from heavy to light primary nuclei
observed by the HiRes collaboration at energies close to 5$\times
10^{17}$ eV~\cite{chem_HiRes}. In this case the UHECR  HiRes
spectrum~\cite{HiRes,hires_mono_spec}, in which the GZK
cutoff~\cite{gzk} is present,   can be fitted with a minimal number of
unknown parameters assuming the extragalactic cosmic rays are only
protons with a power law source spectrum $\sim E^{-\alpha}$ with
$\alpha\simeq 2.6$~\cite{dip}.  This is the minimal UHECR model we
study in this paper.

 Let us mention that the Fermi acceleration 
process predicts lower values of power law index $\alpha\simeq 2.2$. 
which are compatible with the  minimal UHECR model 
if a power law distribution 
of the  maximum  source energies is assumed~\cite{Kachelriess:2005xh}. 
Here we do not study this possibility.

The GZK process produces pions. From the decay of $\pi^{\pm}$ one
obtains neutrinos, the  ``cosmogenic neutrinos"~\cite{bere}. From the
decay of $\pi^0$ we obtain photons,  which we call ``GZK photons".
Previously we studied in detail the GZK photon flux dependence on
different unknown parameters of the assumed source spectrum and
distribution and the intervening cosmological
backgrounds~\cite{gzk_photon}. Here we discuss the perspectives for
photon detection in the minimal UHECR model. Our previous results of
Ref.~\cite{gzk_photon} show that because of the relatively hard source
spectrum required, with $\alpha\simeq 2.6$, the GZK photons are
subdominant at all energies.

The plan of the paper is the  following. In the next section we explain our
calculations.  In Section  \ref{sec:protons} we study the range of parameters  of the
source spectrum  and distribution that best fit the UHECR spectrum,
which is dominated by proton primaries at all energies.  In Section  \ref{sec:photons}
we  show the maximum and minimum expected  level of GZK photons and
comment briefly on cosmogenic neutrinos.  Our conclusions follow in
Section \ref{sec:conclusion}.

\section{Propagation of protons and photons}
\label{sec:propag}

We use a numerical code originally developed in Ref.~\cite{kks1999} to
compute the flux of GZK photons produced by a homogeneous
distribution of sources emitting originally only protons. This is the
same numerical code as in Ref.~\cite{gzk_photon}, with a few
modifications.  This code was compared at the individual reaction
level with the code developed by G.~Sigl and
S.~Lee~\cite{propagLeeSigl} and was already used in several studies of
cosmic ray and secondary gamma-ray and neutrino fluxes~\cite{neutrino}.

 The code uses the  kinematic equation approach and calculates the
 propagation of nuclei, nucleons, stable leptons  and photons using
 the standard dominant  processes. For nucleons, it takes into account
 single and multiple pion  production and $e^{\pm}$ pair production on
 the CMB, infrared/optical and radio  backgrounds, neutron
 $\beta$-decays and the expansion of the Universe. The  hadronic
 interactions of nucleons are now derived from the well established
 SOPHIA event generator~\cite{SOPHIA}, more accurate in the multi-pion
 regime  than the old code in Ref.~\cite{kks1999}. For photons, the
 code includes  $e^{\pm}$ pair production, $\gamma + \gamma_B
 \rightarrow e^+ e^-$, double  $e^{\pm}$ pair production $\gamma +
 \gamma_B \rightarrow e^+ e^-  e^+ e^- $,
 processes. For electrons and positrons, it takes into account inverse Compton
 scattering, $e^\pm + \gamma_B \rightarrow e^\pm \gamma$,
triple pair production,  $e^\pm + \gamma_B \rightarrow
 e^\pm e^+  e^- $ ,  and synchrotron energy loss on extra galactic
 magnetic fields (EGMF).  All these reactions are  discussed in detail
 for example in the Ph.D. thesis of  S.Lee \cite{propagLeeSigl} and  that of
 O.Kalashev~\cite{kks1999}. The propagation of  nucleons and
 electron-photon cascades is calculated self-consistently. Namely,
 secondary (and higher generation) particles arising in all reactions
 are  propagated alongside the primaries.

As it is usual, we take the spectrum of an individual UHECR source  to
be of  the form:
\be  F(E) = \frac{f}{E^\alpha} ~~\Theta (E_{\rm max} -E)
 \label{proton_flux}
\ee
where $f$ provides the flux normalization, $\alpha$ is spectral  index and
$E_{\rm max}$ is the maximum energy to which protons can be
accelerated at the source.

 We assume a standard cosmological model with  a Hubble constant
$H=70$~km~s$^{-1}$~Mpc$^{-1}$, a dark energy density (in units of the
critical density) $\Omega_{\Lambda}= 0.7$ and a dark matter density
$\Omega_{\rm m}=0.3$. The total source density in this model can be
defined by
\be n(z) = n_0 (1+z)^{3+m}  \Theta (z_{\max}-z) \theta (z-z_{\min}) \,,
\label{sources}
\ee
where $m$  parameterizes the source density evolution, in such a   way
that $m=0$ corresponds to non-evolving sources with constant density
per comoving volume,  and $z_{\min}$ and $z_{\max}$ are respectively
the redshifts of the closest and most distant sources.  Sources in the
range $2<z<z_{\max}$ have a negligible contribution  to the UHECR flux
above $10^{18}$~eV. The value of $z_{\min}$ is connected to the
density of sources and influences strongly the shape of  the ``bump"
produced by the pile-up of protons which loose energy  in the GZK
cutoff and the strength of the GZK
suppression~\cite{kst2003,blasi2003,sources2004}.  In the following we
fix $z_{\rm max} =3$ and consider three values for $z_{\rm min}$,
namely  0,  0.005 and  0.01 in Eq.~(\ref{sources}).

The main energy loss mechanism for photons with $E>10^{19}$~eV is
pair production on the radio background (at lower energies pair
production on the CMB is more important), possibly followed by synchrotron
radiation of electrons and positrons. To take into account the effect of the intervening
backgrounds, here we  fit the UHECR data assuming either  minimal
intervening  radio   background (which we take to be the radio
background of Clark {\it et al.}~\cite{clark}) and extragalactic
magnetic field EGMF $B=10^{-11}$ G or a maximal intervening  background
(for which we take the largest radio background of Protheroe and
Biermann~\cite{PB}) and a EGMF $B=10^{-9}$ G, with many different
source models.  A difference with respect  to older versions of our
code is in the infrared/optical background assumed. We use now the
model of Ref.~\cite{Stecker:2005qs}. In any event,  this background is
not very important for the production and absorption of GZK photons at
high energies.

We consider then many different spectra resulting from changing   the
slope $\alpha$ and the maximum energy $E_{\max}$  in
Eq.~\ref{proton_flux} within the ranges $2.3 \leq \alpha\leq 2.9$ and
$1.6 \times 10^{20} {\rm eV}\leq E_{\max}\leq 1.28 \times 10^{21}$~eV
and the source evolution parameter $m$ in Eq.(\ref{sources})  within
the range $-2 \leq m \leq 3$. Notice that  $E_{\max}$ cannot be
smaller than the largest event energy, $1.6 \times 10^{20}$~eV,
observed by HiRes.  We change these parameters in steps
$\alpha_n=2.3+0.05 n$, with $n=1$ to 12, $E_{max-\ell}=1.6 \times
10^{20}eV \times 2^\ell$, with $\ell=0$ to 6. and $m_i=-2+ i$ with
$i=0$ to 5.

 For each one of the models so obtained we compute the  predicted
 UHECR spectrum  by summing up the contributions of protons plus GZK
 photons arriving to us from  all sources. In general models we need
 to consider a larger range of spectral indeces~\cite{gzk_photon2}. In
 the minimal UHECR model we study here instead, one  fits the observed
 spectrum UHECR down to energies $E=1-2$ EeV with  extragalactic
 protons, which requires a steaply falling source proton spectra  with
 $\alpha \ge 2.3$. For such injected  proton spectra the GZK photons
 reaching us are subdominant at all energies. So, fitting the observed
 HiRes  data with the sum of protons and photons arriving to us from
 all sources is  almost equivalent to using just the protons reaching
 us.
 
With the spectrum predicted for each combination of parameters we fit
the UHECR data from  2 $\times 10^{18}$~eV up to the end point of the
HiRes spectrum (i.e. the 28  highest  energy   bins of the HiRes 1 and
2 combined monocular data) plus one extra bin at energies above the
``end point'' (the point in energy beyond which no events were
observed)  of the spectrum.   This last additional bin with zero
observed events, extends from the end point of the observed spectrum
to the maximum energy $E_{\max}$ assumed for the  injected spectrum in
Eq.~\ref{proton_flux}. This extra bin takes into account   the
non-observation of events above the highest occupied energy bin in the
data HiRes, i.e. at $E> 1.6 \times 10^{20}$~eV for the HiRes spectrum
we used~\cite{hires_mono_spec}). We compute the expected number of
events in this last bin using an exposure that we derive from the
HiRes data above 10$^{20}$~eV, by comparing the published integrated
fluxes with the number of events observed and assuming the exposure is
energy independent (above 10$^{20}$~eV).

 To fit the UHECR data with each predicted spectrum we  follow a
 procedure  similar to that of Ref.~\cite{Fodor-K-R} applied to the
 bins just mentioned. We reconstruct the measured number of events in
 each bin from the published data of  HiRes (using the error
 bars~\cite{Poisson-errors}) and compare them with the number of
 events in each bin predicted by each of the models.   We choose the
 value of the parameter $f$ in Eq.~\ref{proton_flux}, i.e. the
 amplitude of the injected spectrum, by maximizing the Poisson
 likelihood function, which is equivalent to minimizing $-2
 \ln{\lambda}$, (i.e. the negative of the log likelihood
 ratio)~\cite{statistics}. This procedure amounts to choosing the
 value of $f$  so that the mean total number of events predicted
 (i.e. the sum of the average predicted number of events in all fitted
 bins)  is equal to the total number of events observed. We then
 compute using a Monte Carlo technique the goodness of the fit, or
 $p$-value of the distribution, defined as the mean fraction of
 hypothetical experiments (observed spectra)  with the same fixed
 total number of events, which would result in a worse, namely lower,
 Poisson likelihood than the one obtained (in the maximization
 procedure that fixed $f$). These hypothetical experiments are chosen
 at random according to a multinomial distribution. We have checked
 that this procedure  when applied to bins with large number of events
 gives the same results as  a Pearson's $\chi^2$ fit, both for the
 value of the normalization parameter $f$ and for the goodness of fit.
 A higher $p$ value corresponds to a better fit, since more
 hypothetical experimental results would yield a worse fit than the
 one we obtained.  We make one additional requirement on the fit that
 insures  that the predicted flux does not exceed  the observed flux
 at energies below 2 $\times 10^{18}$eV.

In the next section we present our results the for total UHECR flux, which is
dominated  by protons at all energies.

\section{The proton flux}
\label{sec:protons}

In this section we  find the range of source proton spectrum and
distribution parameters $\alpha$, $E_{max}$, $z_{min}$ and  $m$,
consistent with the HiRes observed spectrum~\cite{HiRes} at  energies
$E \ge 2$ EeV.

\begin{figure}[ht]
\includegraphics[height=0.5\textwidth,clip=true,angle=270]{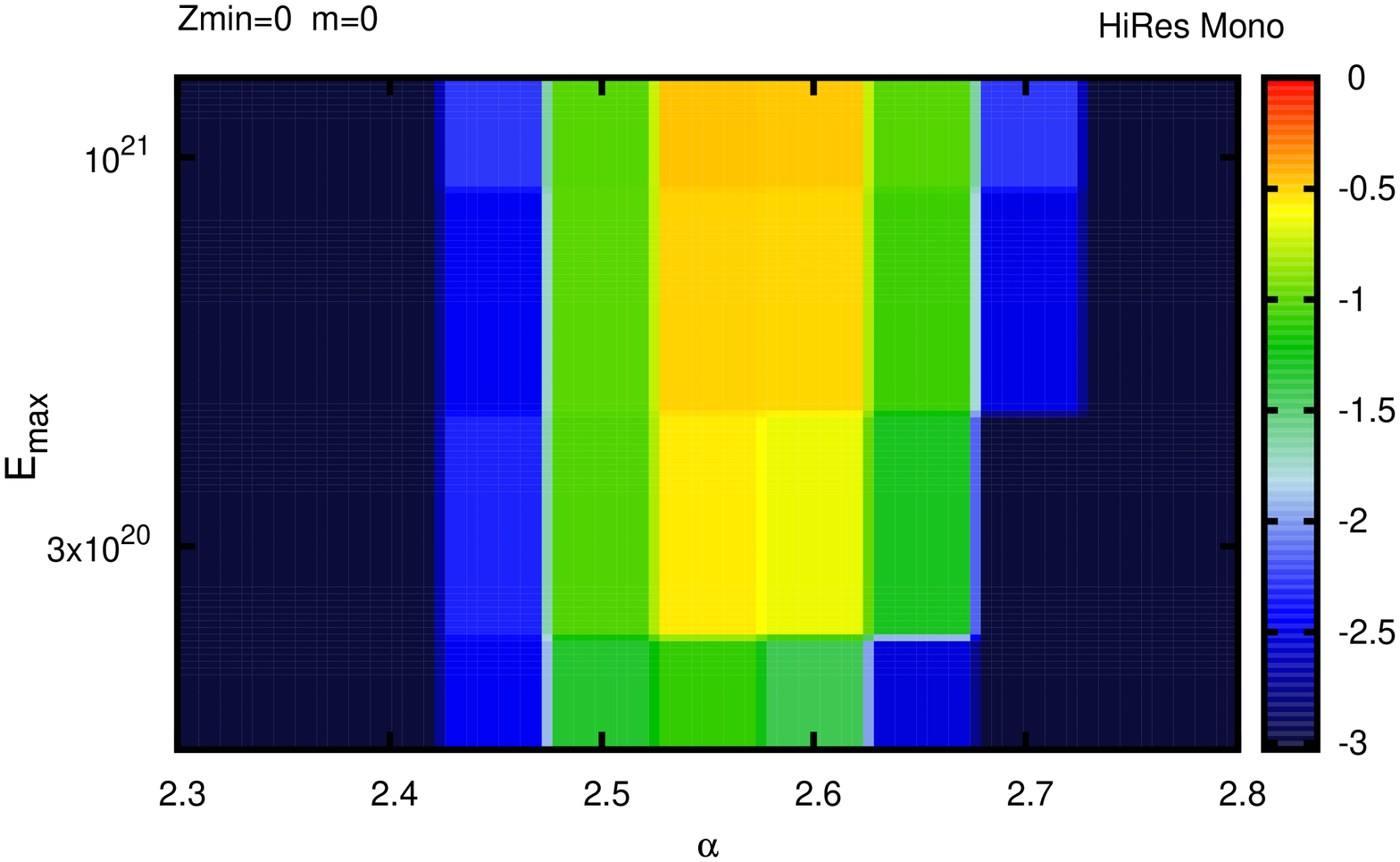}
\includegraphics[height=0.5\textwidth,clip=true,angle=270]{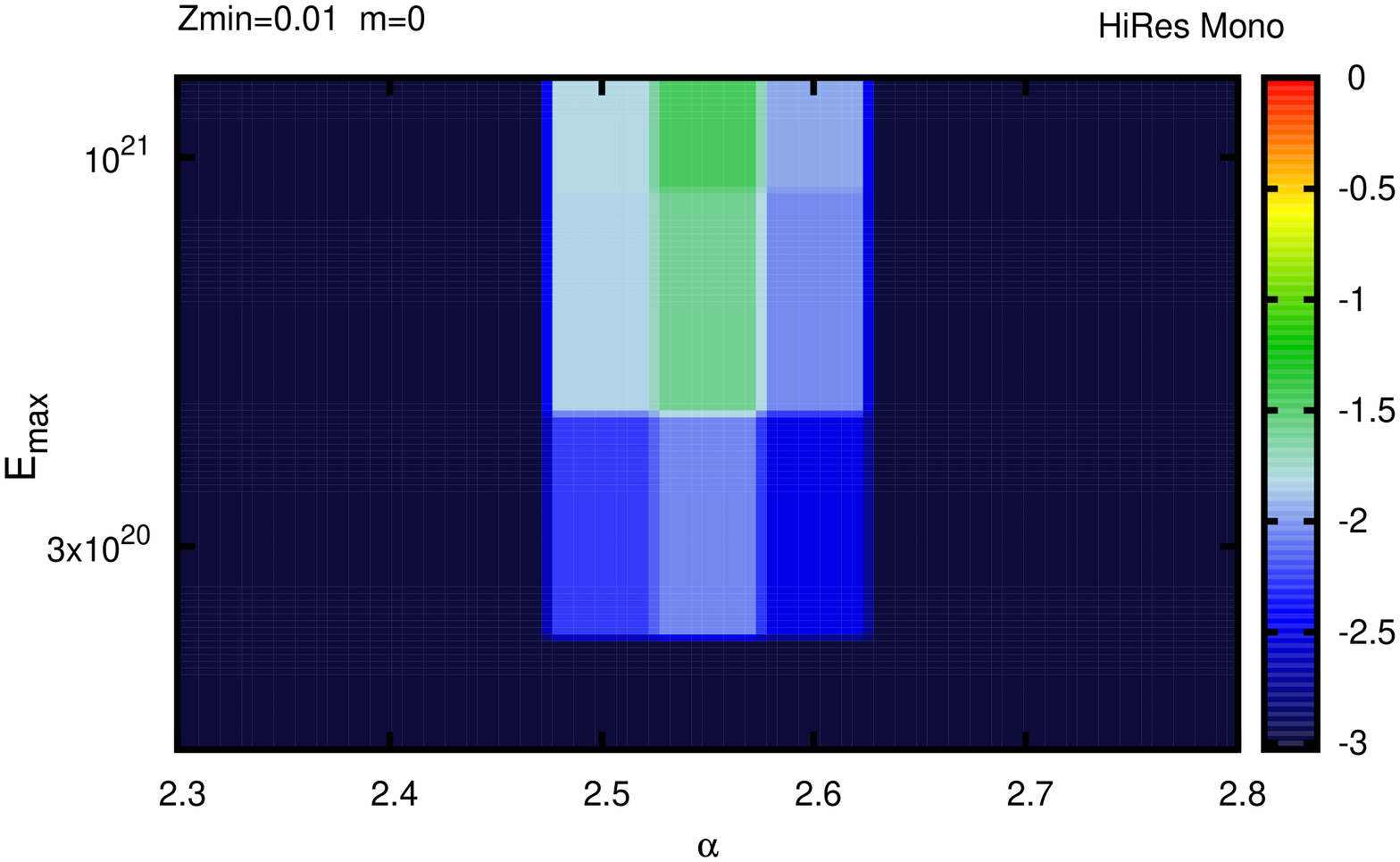}
\caption[...]{Consistency level of the predicted UHECR proton flux
with HiRes data as function of  $E_{max}$ and $\alpha$  for $m=0$ and
either  $z_{min}=0$ (in Fig.~\ref{Fig_proton_E_alpha}a, left panel),
i.e. a continuous distribution of sources, or $z_{min}=0.01$
(Fig.~\ref{Fig_proton_E_alpha}b, right panel), i.e. with no sources
within a 50 Mpc radius.  Color coded logarithmic $p$-value scale, from
best ($p=1$) to worse ($p$ close to zero).  }
\label{Fig_proton_E_alpha}
\end{figure}

\begin{figure}[ht]
\includegraphics[height=0.5\textwidth,clip=true,angle=270]{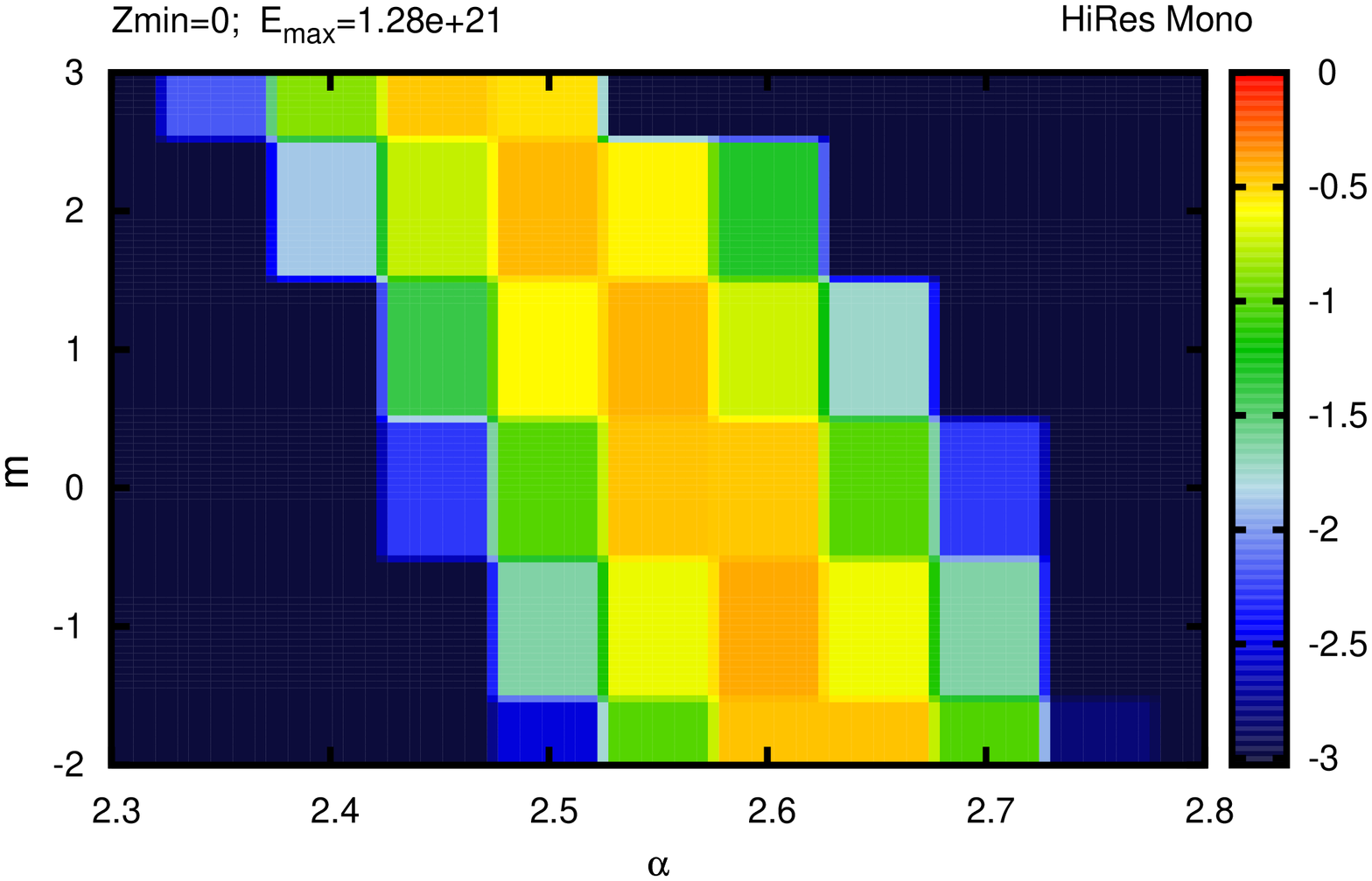}
\includegraphics[height=0.5\textwidth,clip=true,angle=270]{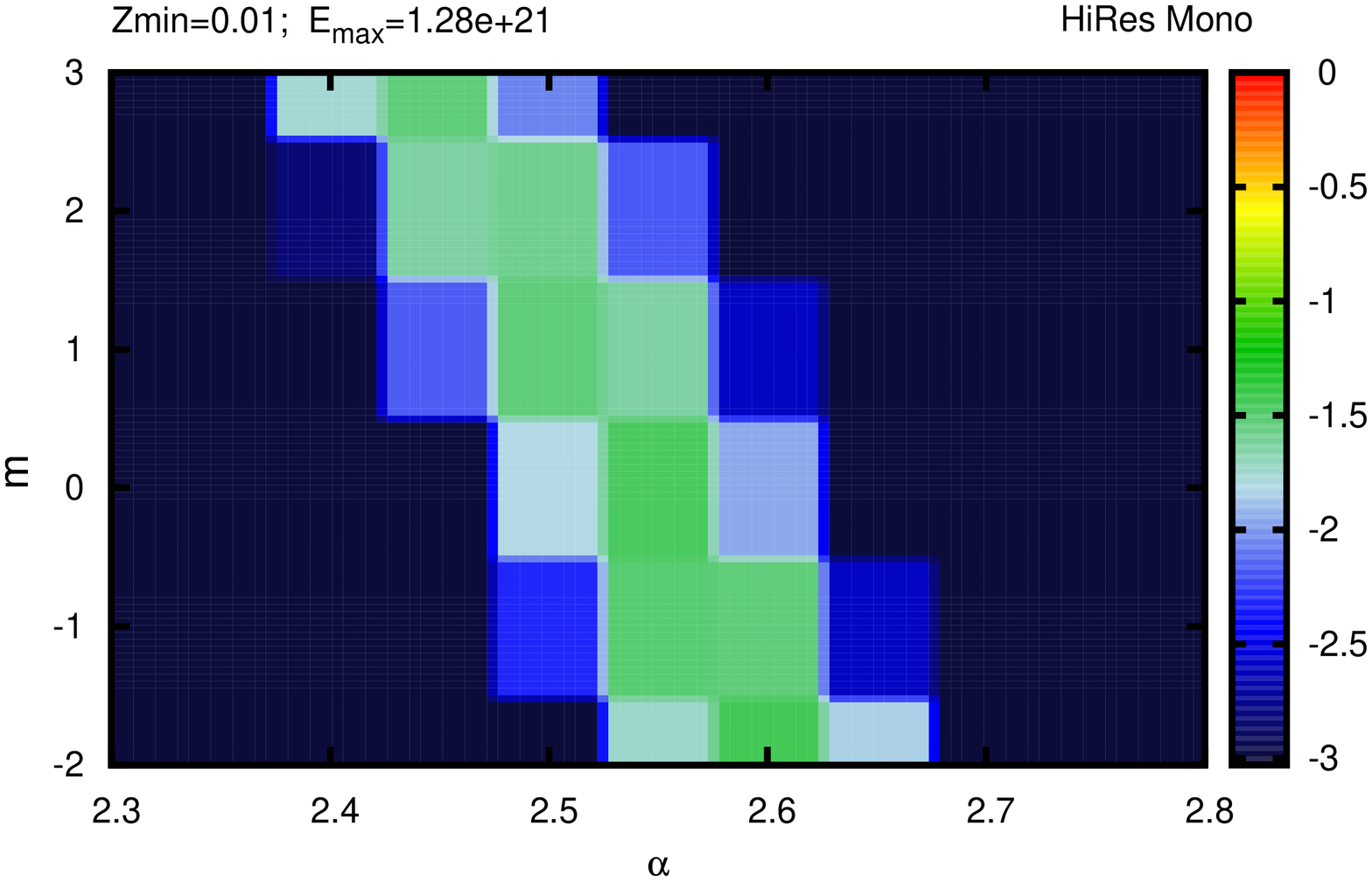}
\caption[...]{Consistency level of the predicted UHECR proton flux
with HiRes data as function of  $m$ and $\alpha$ for $E_{max}=10^{21}$
eV and   either  $z_{min}=0$ (in Fig.~\ref{Fig_proton_m_alpha}a, left
panel),  i.e. a continuous distribution of sources, or $z_{min}=0.01$
(Fig.~\ref{Fig_proton_m_alpha}b, right panel), i.e. with no sources
within a 50 Mpc radius.  Color coded logarithmic $p$-value scale, from
best ($p=1$) to worse ($p$ close to zero).  }
\label{Fig_proton_m_alpha}
\end{figure}

In Fig.~\ref{Fig_proton_E_alpha}  and  Fig.~\ref{Fig_proton_m_alpha}
we show the logarithm of the $p$-value in a color coded scale, from
best ($p=1$) to worse ($p$ close to zero), which measures the
consistency level of the predicted UHECR proton flux  with the HiRes
data, for different parameter ranges.

\begin{figure}[ht]
\begin{center}
\includegraphics[height=0.8\textwidth,clip=true,angle=270]{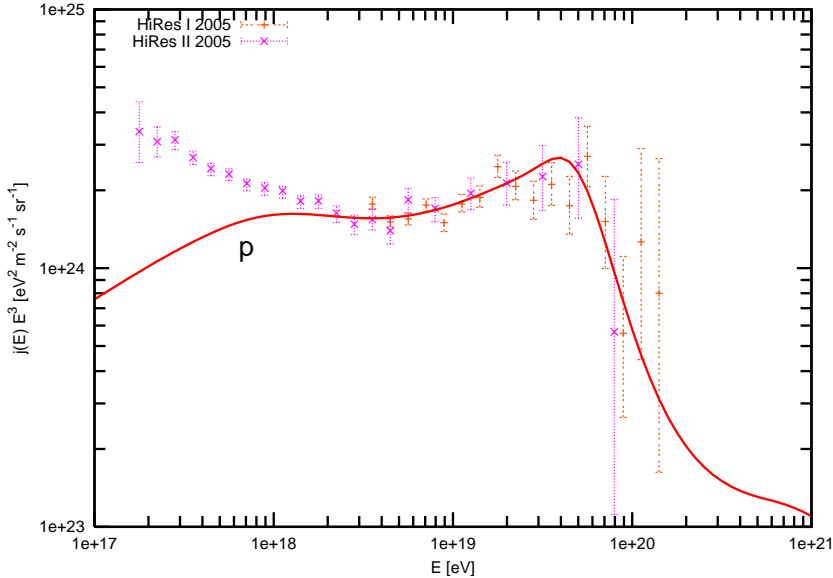}
\end{center}
\caption[...]{UHECR proton flux fitted to the HiRes data in the energy
range  $E>2\times 10^{18}$~eV of one of the models with best goodness
of fit:  $m=0$, $z_{min}=0$,  $E_{max}=10^{21}$ eV  and $\alpha=2.55$.
}
\label{Fig_proton_spectrum}
\end{figure}

The high energy  part of the predicted  spectrum depends mostly on the
power law index $\alpha$, the maximum injected proton energy $E_{max}$
and the minimal distance  to  the sources $z_{min}$. In
Fig.~\ref{Fig_proton_E_alpha} the $p$-values are shown as function of
$E_{max}$ and $\alpha$ for $m=0$ and either $z_{min}=0$
(Fig.~\ref{Fig_proton_E_alpha}a, left panel), i.e. a continuous
distribution of sources,  or $z_{min}=0.01$
(Fig.~\ref{Fig_proton_E_alpha}b, right panel), i.e. with no sources
within a 50 Mpc radius.  We can see in the figure that  fitting the
UHECR data from  $2$ EeV and above, the initial proton spectrum should
be relatively hard, with $\alpha=2.50-2.65$ in
Eq.(\ref{proton_flux}). Fig.~\ref{Fig_proton_E_alpha}a shows that this
range does not depend strongly on  $E_{max}$ for a continuous
distribution of sources. If there are no sources within a distance of
50 Mpc distance,  thus $z_{min}=0.01$, as shown
inFig.~\ref{Fig_proton_E_alpha}b, the HiRes observed spectrum is not
fitted  as well anymore, and a relatively high maximum energy $E =
10^{21}$ eV is required for a reasonable fit,  with, say,  $p >0.05$.

The low energy  part of the predicted  spectrum depends mostly on the
power law index $\alpha$ and source evolution index $m$.  In
Fig.~\ref{Fig_proton_m_alpha} we show the goodness of fit $p$-value  as function
of  $m$ and $\alpha$ for $E_{max}=10^{21}$ eV and again for either
$z_{min}=0$ (Fig.~\ref{Fig_proton_m_alpha}a, left panel) or
$z_{min}=0.01$ (Fig.~\ref{Fig_proton_m_alpha}b, right panel). This
figure cleary shows the degeneracy between the  parameters $m$ and
$\alpha$: as $m$ increases from $-2$ to 3 the value of $\alpha$ of the
best fits decreases from $\simeq 2.6-2.7$ to $\simeq 2.4-2.5$. Again
the fit is worse for $z_{min}=0.01$, in which case the $p$-value is
never higher than 0.04.

Fig.~\ref{Fig_proton_spectrum} shows the total predicted UHECR
spectrum fitted to the HiRes data in the energy range  $E>2\times
10^{18}$~eV of one of the models with best goodness of fit, as can be
seen in Figs.~\ref{Fig_proton_E_alpha}a and \ref{Fig_proton_m_alpha}a. It has $m=0$, $z_{min}=0$,
$E_{max}=10^{21}$ eV and $\alpha=2.55$.

\section{The GZK photon flux}
\label{sec:photons}

In this section we discuss the secondary photon fluxes. The main
difference between  the minimal model  we are concentrating on here and
other models (see Ref.~\cite{gzk_photon}) is that in the minimal model
one fits the UHECR data with  extragalactic protons data starting from
low  energies $E> 2$ EeV, what requires  a hard
spectrum with  index $\alpha>2.4$ 
(see Figs.~\ref{Fig_proton_E_alpha} and \ref{Fig_proton_m_alpha}).
In this case the GZK photon flux is always sub-dominant, at all
energies.

\begin{figure}[ht]
\begin{center}
\includegraphics[height=0.8\textwidth,clip=true,angle=270]{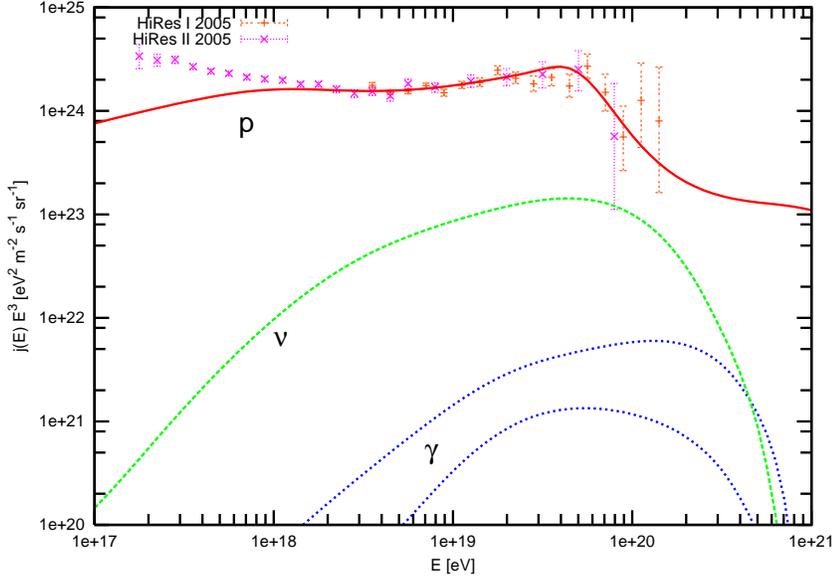}
\end{center}
\caption[...]{GZK photon and  cosmogenic neutrino spectra (besides the
proton spectrum) for the model of Fig.~\ref{Fig_proton_spectrum}
($m=0$, $z_{min}=0$,  $E_{max}=10^{21}$ eV  and $\alpha=2.55$).  The
upper photon line is for minimal radio background and
$B_{EGMF}=10^{-11}$ G,  while the lower photon line for maximal  radio
background $B_{EGMF}=10^{-9}$ G.}
\label{Fig_photon_spectrum}
\end{figure}

As an example, in   Fig.~\ref{Fig_photon_spectrum}  we show the
possible range of GZK photon fluxes for the predicted proton spectrum
of Fig.~\ref{Fig_proton_spectrum}. The range of photon fluxes is
between the upper photon (blue-dotted) line which was calculated with
minimal radio background and $B_{EGMF}=10^{-11}$~G and the lower
photon line corresponding to maximal radio background  and
$B_{EGMF}=10^{-9}$~G. (How  the GZK photon flux depends on the radio
background and extragalactic magnetic fields assumed can be  seen in
Ref.~\cite{gzk_photon}).

Here we do not deal with neutrinos in any detail, but just to compare
the photon and neutrino fluxes produced in the same GZK processes, in
Fig.~\ref{Fig_photon_spectrum} we also plotted the cosmogenic neutrino
flux per flavor for the same model.  Even if the neutrino flux is much
higher than the photon flux, its detection may be even more difficult
due to the strongly reduced probability of neutrinos to produce
air-showers.

In Fig.~\ref{Fig_photon_spectrum} one can see that the best energy
range to find  GZK photons is $E=5-20$ EeV.  At higher energies,  the
small event statistics  will not  allow to find a 1\% fraction of
photons in the UHECR flux, while  at lower energies the photon fraction is
 strongly reduced.

\begin{figure}[ht]
\includegraphics[height=0.5\textwidth,clip=true,angle=270]{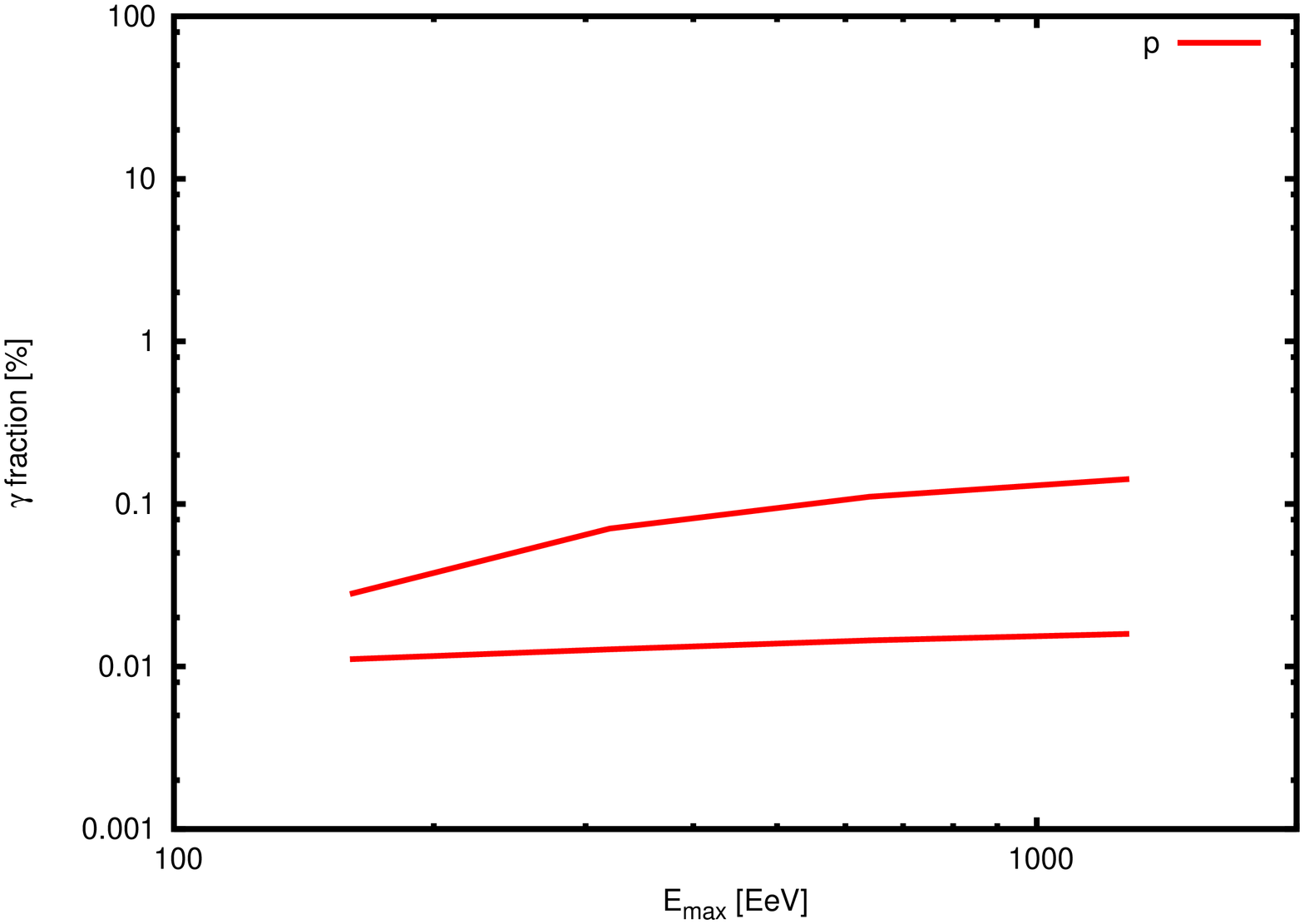}
\includegraphics[height=0.5\textwidth,clip=true,angle=270]{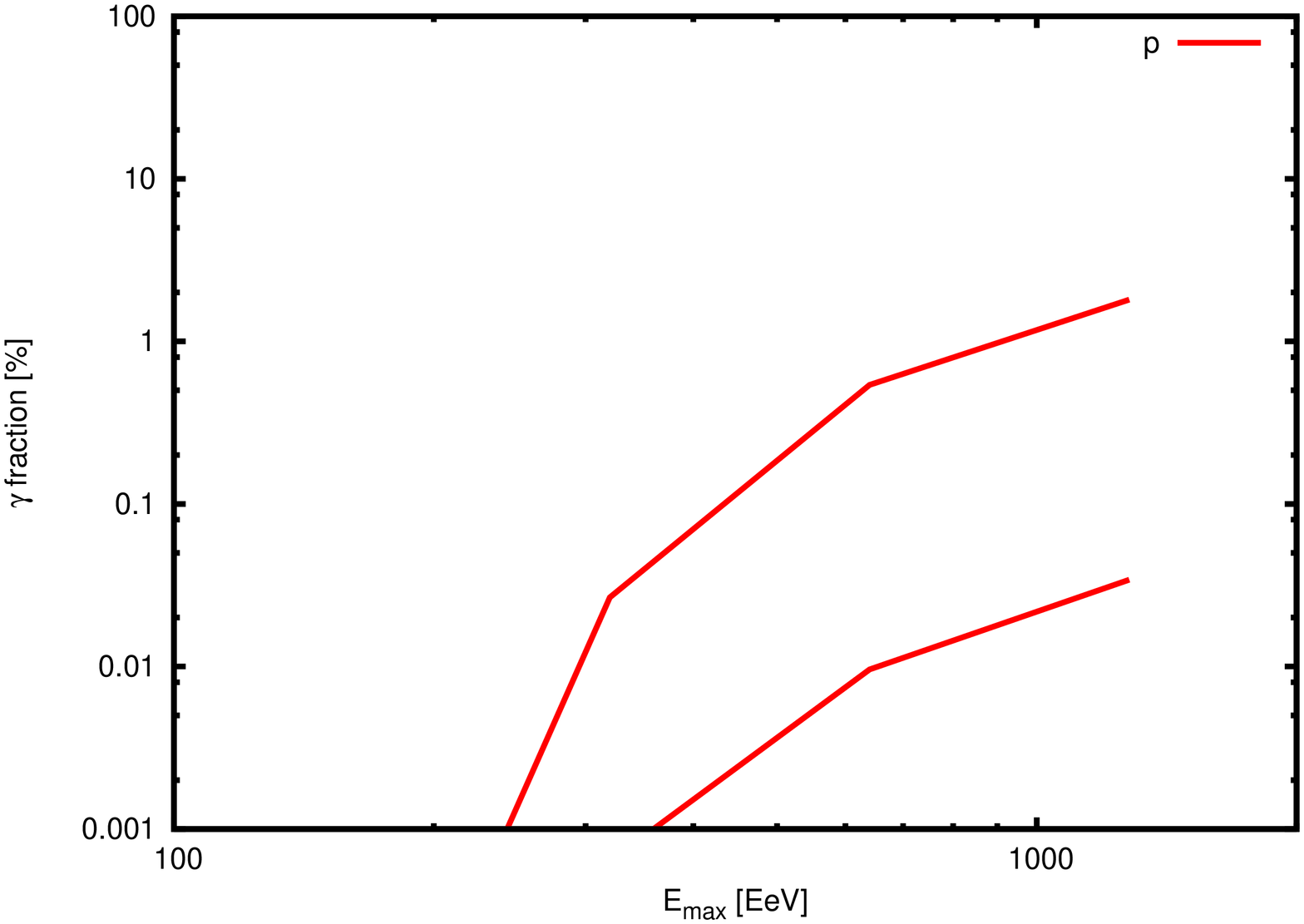}
\caption[...]{Maximum and minimum GZK photon fractions  given  in
percentage  of the integrated fluxes above  the energy $E$ as function
of $E_{max}$, the maximum energy of source proton spectra,  for
$E=10^{19}$ eV (Fig.~\ref{Fig_photon_spectrum}a, left panel)  and $E=
10^{20}$ eV (Fig.~\ref{Fig_photon_spectrum}b, right panel).  }
\label{Fig_photon_Emax}
\end{figure}

The dependence of the GZK photon fractions  on $E_{max}$, the maximum
source proton energy,    is shown  in Fig.~\ref{Fig_photon_Emax}. The
figure shows  the maximum and minimum GZK photon fractions  given  in
percentage of the integrated UHECR fluxes above  the energy $E$  for
$E=10^{19}$ eV (left panel) and $E= 10^{20}$ eV (right panel).  In
order to define the range of possible photon fluxes we use only models
with p-values $p>0.05$ (i.e. we eliminate those models which are
inconsistent  with the HiRes observed spectrum at the 95 \% C.L.). 
Those shown are the maximum and minimum GZK photon fractions obtained
for each value of $E_{max}$ by varying all the other parameters as
specified above, and choosing either minimum or maximum intervening
backgrounds.  The important conclusion coming from this figure is the
stability of the GZK photon fractions at  $E>10^{19}$ eV as function
of $E_{max}$. Notice that the expected photon fractions are between
10$^{-4}$ and 10$^{-3}$ for the whole $E_{max}$ range we consider (see
the left panel).  This contrasts with the situation at $E>10^{20}$ eV
(right panel), where the photon fractions depends strongly on
$E_{max}$.

\begin{figure}[ht]
\begin{center}
\includegraphics[height=0.8\textwidth,clip=true,angle=270]{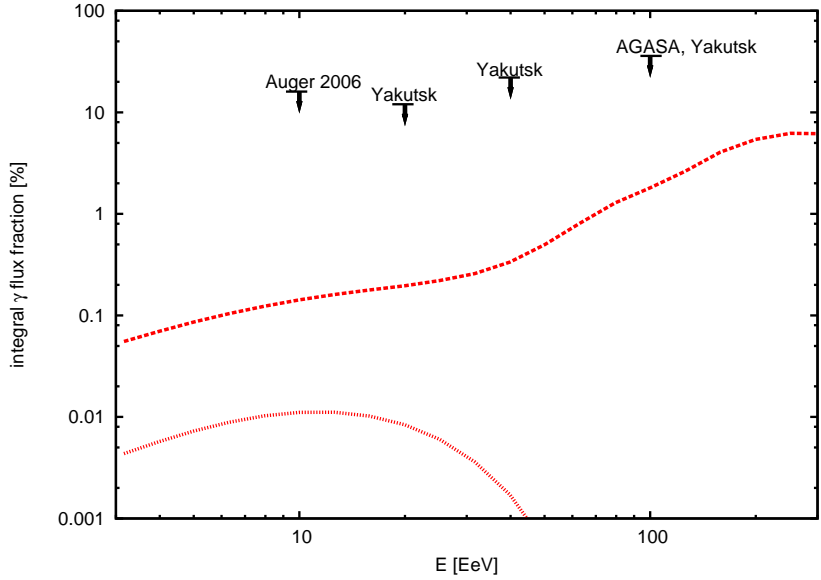}
\end{center}
\caption[...]{Maximum and minimum GZK photon fractions  given  in
percentages  of the integrated flux above  the energy $E$ as function
$E$ for maximum source proton energy $E_{max}=10^{21}$ eV. Present
limits on photon  fraction from Auger~\cite{AugerLimit},
Yakutsk~\cite{Yakutsk_2007} and combined AGASA/Yakutsk
\cite{AgasaYakutskLimit}  data are also shown.}
\label{Fig_photon_E_cut}
\end{figure}

Finally, in Fig.~\ref{Fig_photon_E_cut} we show the GZK photon
fraction given in percentage of the integrated UHECR flux above  the
energy $E$ as function of $E$,  for the whole parameter space we consider
(i.e. maximum source proton energy $1.6 \times 10^{20} eV \leq
E_{max}\leq 1.28 \times 10^{21}$ eV,  source evolution parameter  $-2
\leq m \leq 3$, power law index  $2.3 \leq \alpha \leq 2.9$ and
minimum redshift of the sources $0 \leq z_{min} \leq 0.01$).  Present
limits on the  photon fraction from Auger~\cite{AugerLimit},
Yakutsk~\cite{Yakutsk_2007} and combined AGASA/Yakutsk~\cite{AgasaYakutskLimit} data are also shown in the figure.  It is
clear that, contrary to the case of top-down models (which are
restricted already by present bounds on the GZK-photon
fraction~\cite{gzk_photon}) the  present limits are well above the
expected the GZK photon fraction in the minimal UHECR model by a
factor  of 10  to 100 depending on the energy (see
Fig.~\ref{Fig_photon_E_cut}). The detection of GZK photons in this
model will remain as a task for the future.

We have also checked the dependence of  the photon fractions in Fig.~\ref{Fig_photon_E_cut}
on the  lowest energy for which we fit the HiRes data by changing this energy in the interval $0.3 - 2$ EeV. We  found no significant changes with respect to  Fig.~\ref{Fig_photon_E_cut} at energies close to
 $E=10$ EeV, and small changes close to $100$ EeV.
The highest photon fractions are always at the highest energies,  but the best energies
to observe photons are close to $10$ EeV due to the
larger  experimental statistics as well as the smaller dependence of our predictions on  the unknown parameters at these energies.

Thus, the expected  photon fraction of the integrated flux above
$E=10$ EeV in the minimal UHECR models, is $10^{-4}$ to $10^{-3}$ independently of the unknown
parameters we considered. Already Auger South after 5 years of data
taking can  reach a statistics of $10^4$ events at energies  $E>10$
EeV. This would allow, in principle, to detect GZK photons, if they
can be discriminated from the large background of proton cosmic rays.

\section{Discussion and conclusions}
\label{sec:conclusion}

The South site of the Pierre Auger Observatory after several years of
data taking will probably be able to reach a photon fraction
sensitivity of the order of $10^{-3}$ in the integrated flux close to
$E=10$ EeV.  As can be seen in Fig.~\ref{Fig_photon_E_cut} this is the
level of the largest GZK photon fraction expected in the minimal UHECR
model. Larger future observatories like Auger North plus South \cite{Auger_North} and EUSO~\cite{EUSO} could probe lower photon fractions if they
are able to collect statistics a  factor of 5-10 larger 
than Auger South and have  thresholds around $1-2 \times 10^{19}$ eV
(provided these experiments
are sensitive to photon primaries).

We have assumed that  the sources emit only protons, however our predictions
for GZK photon fractions  shown in Fig.~\ref{Fig_photon_E_cut}  would
not change  too much if nuclei primaries were present too,  as assumed
in  the so called ``mixed models"~\cite{mixed_model}. The reason is
that even in mixed models, primary protons dominate the UHECR flux
at high energies  $E>50$ EeV, i.e. in the energy  region where the primary
protons produce secondary GZK photons.

 Let us also mention that the photon flux at high energies $E>10^{18}$
 eV  could be enhanced by the interaction of UHECR protons with
 energies $E \sim 1-4 \times 10^{19}$ eV with a large infrared
 background  in our galaxy~\cite{Tinyakov:2006jm} due to the reaction
 $p+\gamma_{IR} \rightarrow p + \pi^0 \rightarrow p+ 2 \gamma$.  
  Only $2 \times 10^{-4}$ of the UHECR protons with energy 10-50 EeV would
 interact with the infrared background  in our galaxy. Thus,   the
 resulting photons, which would have an energy 1-5 EeV, would
 constitute a small fraction  of the order of $ 4\times 10^{-6}$ of the 
 integrated UHECR flux at these energies.
 These are much smaller than the expected GZK
 photons at energies 1-5 EeV (as shown in Fig.~\ref{Fig_photon_E_cut}).

As a final remark let us mention that even if the GZK photon fluxes
considered here are very small, much larger fluxes are possible in
more general models, which are not restricted  by the condition that
all the UHECR spectrum from energies $2\times 10^{18}$ eV to the
largest is explained  with extragalactic protons~\cite{gzk_photon}.

In conclusion, here we systematically study the possible GZK photon fluxes
in the minimal UHECR model in the multi-dimentional
parameter space of source proton spectrum power law index $\alpha$ and
maximum energy $E_{max}$, minimal
distance to the sources
$z_{min}$ (which is directly connected to the source density), source
evolution parameter $m$,
average magnetic field  value $B$ and extragalactic radio background in
the interval $10^{-11} G  \le B \le 10^{-9} G$.
We also consider the dependence of our results on the lowest energy of the
HiRes spectrum
we chose to fit, varying it in the  interval
$3 \times 10^{17} ~{\rm eV}~ \le E_{c} \le 2 \times 10^{18}$ eV.
In each case we take into account only the models which are consistent with
spectrum of cosmic rays observed by HiRes experiment at the 95 \% C.L.
Our results,
presented in Fig.~\ref{Fig_photon_E_cut}, show that future experiments have
to reach a
sensitivity of $10^{-4} - 10^{-3}$ in the photon to proton fraction
for energies $E \ge
10^{19}$ eV
in order to detect GZK photons in the minimal UHECR model.

Finally, we want to mention that after this work was finished we got the
draft of a paper by G.~Sigl in which the GZK photon
flux in the minimal UHECR model is also studied. In this paper G.~Sigl
mainly investigates the effect on the photon fluxes of a three-dimentional
magnetic field stucture, and gives only examples of possible photon fluxes in
several cases. Here, instead, as mentioned above,  we simplify the  effect
of extragalactic
magnetic fields, considering only  their average value  to be in the
interval $10^{-11} G  \le B \le 10^{-9} G$ , while
we systematically study the possible photon fluxes in their multi-dimentional
parameter space.

\vspace{0.3cm}
{\bf Acknowledgments}

The work of G.G and O.K.
 was supported in part by NASA grant NAG5-13399.
G.G was also supported in part by the US DOE grant DE-FG03-91ER40662
Task C.

%
%
%
%

\end{document}